\documentclass[
    ,final            % use final for the camera ready runs
  ]
  {aipproc}

\layoutstyle{8x11double}

\def\sun{$_\odot$}

\begin{document}

\title{Monitoring and Discovering X-ray Pulsars in the Small Magellanic Cloud}

\author{R.H.D. Corbet}{
  address={Laboratory for High Energy Astrophysics,
Code 662, NASA Goddard Space Flight Center,\\ Greenbelt, MD 20771}
  ,altaddress={Universities Space Research Association} % additional visiting address
}

\author{S. Laycock}{
  address={School of Physics and Astronomy, Southampton University,
SO17 1BJ, United Kingdom}
  ,altaddress={Current address:
Harvard-Smithsonian Center for Astrophysics} % additional visiting address
}

\author{M.J. Coe}{
address={School of Physics and Astronomy, Southampton University,
SO17 1BJ, United Kingdom}}

\author{F.E. Marshall}{
  address={Laboratory for High Energy Astrophysics,
Code 662, NASA Goddard Space Flight Center,\\ Greenbelt, MD 20771}
}

\author{C.B. Markwardt}{
  address={Laboratory for High Energy Astrophysics,
 Code 662, NASA Goddard Space Flight Center,\\ Greenbelt, MD 20771}
  ,altaddress={University of Maryland} % additional visiting address
}

\begin{abstract}
Regular monitoring of the SMC with RXTE has
revealed a huge number of X-ray pulsars. Together with discoveries from
other satellites at least 45 SMC pulsars are now known.  One of these
sources, a pulsar with a period of approximately 7.8 seconds, was first
detected in early 2002 and since discovery it has been found to be in
outburst nine times.  The outburst pattern clearly shows a period of 45.1
$\pm$ 0.4 d which is thought to be the orbital period of this system.
Candidate outburst periods have also been obtained for nine other
pulsars and continued monitoring will enable us to confirm these.
This large number of pulsars, all located at approximately the same
distance, enables a wealth of comparative studies.  In addition, the
large number of pulsars found (which vastly exceeds the number expected
simply by scaling the relative mass of the SMC and the Galaxy) reveals
the recent star formation history of the SMC which has been influenced
by encounters with both the LMC and the Galaxy.
\end{abstract}

\maketitle

%%%%%%%%%%%%%%%%%%%%%%%%%%%%%%%%%%%%%%%%%%%%
%% MAINMATTER
%%%%%%%%%%%%%%%%%%%%%%%%%%%%%%%%%%%%%%%%%%%%

\section{The early history of SMC X-ray pulsars}

The first known X-ray pulsar in the SMC was
the persistent supergiant system SMC X-1.
Two luminous transients (SMC X-2, SMC X-3) were discovered with SAS-3.
\cite{C78}. These were thought to be transient Be/neutron star systems
although pulsations were not detected due to the low sensitivity of SAS-3.
It was
hypothesized that SMC pulsars were exceptionally luminous, possibly
related to the low metallicity of the SMC. \cite{Westerlund90}
This was later to be disproved and an alternative explanation
found for the high luminosity of the first few SMC
X-ray pulsars to be discovered. Over the years a few pulsars 
were also found with satellites such 
as ROSAT. \cite{Hughes94}

\section{SMC X-ray pulsars with RXTE}

Serendipitous RXTE slew observations in 1997
showed a possible outburst from the vicinity of SMC X-3.
Follow up target of opportunity
pointed RXTE observation showed a complicated power spectrum with several 
peaks that were not all harmonically related to each other.
Imaging ASCA observations were next made
which showed two separate pulsars, however neither was found to
be located at the position of SMC X-3.
A more
detailed look at RXTE power spectrum showed 
that in fact three pulsars were simultaneously active \cite{C98}.
These observations were the first sign of the existence
of a very large SMC X-ray pulsar
population.

\section{The RXTE monitoring program}

RXTE has been regularly monitoring the SMC since 1997. 
We have discovered very many transient X-ray pulsars. 
For those sources where optical counterparts have been
identified they are all found to be Be stars.
We primarily
make weekly observations of one particularly active region near SMC X-3.
Other SMC regions have been monitored monthly depending on the amount
of time awarded in a particular observing cycle.
We 
use power spectrum to extract pulsed flux from any pulsars. In this
way, although the PCA is not an imaging instrument, the pulsed flux from
multiple sources can be monitored independently.
When new sources are identified we use cross scans in Right Ascension
and declination to localize the position of the new sources. Initially
scans were done as Target of Opportunity observations. However, this
was not always successful if the target had faded before the TOO could be
performed. Therefore we currently include R.A./dec. scans in all our
SMC observations.
For new sources we thus obtain at least minimal positional information.
However, position determination can be problematic if too many pulsars
are simultaneously active.
A log of known X-ray pulsars in the SMC is currently
maintained at  \url{http://http://lheawww.gsfc.nasa.gov/~corbet/pulsars/}.

\section{Orbital periods}
Ten pulsars have candidate orbital periods determined by us from
long term periodicities in their pulsed flux \cite{L04}.
The statistical significance of the
orbital period determinations varies depending on, for example,
the number of outbursts detected and whether the
outbursts are ``type I'' or ``type II'' (i.e. related to periastron passage or not).
In addition, optical monitoring of optical counterparts, has also revealed
the orbital periods of several systems \cite{Cowley03} \cite{Edge04a}.
For those sources where only a small number of outbursts have
been seen so far
continued monitoring should conclusively pin down the orbital periods.
For two SMC X-ray pulsars we have very solid
orbital periods determined from fairly recent
observations.
For the 7.78s pulsar (recently determined to be
SMC X-3 \cite{Edge04b}) we observed 9 outbursts in 2002
and early 2003. All of these outbursts
are consistent with P$_{orb}$ = 45.1 $\pm$ 0.4 days.
More recently, a set of five outbursts from the 144s pulsar
in 2003 and 2004 have revealed an orbital period of 61.4 $\pm$ 1.1 days.

The orbital periods derived so far appear to follow the correlation
between orbital period and spin period found for Galactic Be star systems
(see e.g. Corbet 1986 \cite{C86}). 

\section{Why are there so many pulsars in the SMC?}
A simple scaling based on the 
relative masses  of the SMC and Galaxy predicts only $\sim$2 SMC X-ray pulsars. However
about 45 are now known and are listed in the table.
A more realistic prediction of the expected number of X-ray pulsars
in the SMC should come from the star formation rate as high-mass
X-ray binaries have rather short lifetimes.
However, estimates of 
the current star formation rate in the SMC do {\em not} give very large
numbers. Estimates of the SMC star formation rate
from integrated colors \cite{Rocca81} and from
H$\alpha$ \cite{Kennicutt94} give 0.064 and 0.046 M\sun/yr respectively
compared to the Galactic star formation rate of a few M\sun/yr. 
However, SMC has experienced encounters with LMC and Galaxy in past
and these
encounters likely triggered bursts of star formation (see e.g.
Yoshizawa \& Noguchi 2003 \cite{Yoshizawa03}).
The remnants of this star formation burst may now seen
as high mass Be star X-ray binaries.

\section{The potential of the SMC pulsar database}
All the SMC pulsars lie at approximately same distance (although
the SMC does have a substantial depth).
This facilitates comparative studies (e.g. pulse profiles) as function of luminosity.
We are continuing to
build up a huge X-ray pulsar database with RXTE 
which can be used to study pulsar parameters in may different ways.
For example, we can now start to compare the pulse period distributions
and the pulse/orbital diagrams for the SMC and the Galaxy.
One notable difference so far between the SMC and the Galaxy is the apparent
lack of supergiant wind accretion systems in the SMC. In the Galaxy these
sources have luminosities of a few 10$^{36}$ to 10$^{37}$ ergs/s and,
unless the SMC's lower metallicity causes such systems to have very much
lower luminosities,
should be detectable in our observations which have limiting luminosities
of roughly 10$^{36}$ ergs/s. Although our weekly monitoring observations
have durations of about 7000s, which makes them somewhat less sensitive
to the longer pulse periods exhibited by wind accretion systems, we have
had some longer duration observations and long Chandra and XMM observations
should also have been sensitive to detection of this type of system.
The overall pulsar properties of the SMC can tell us about 
the relative evolution 
of a very nearby galaxy and we can compare source count properties
with predictions of various models.

\section{SMC pulsar monitoring beyond RXTE}
Although Chandra and XMM are sensitive to source detection, 
neither is well suited to frequent monitoring to determine orbital periods.
The RXTE ASM is just sensitive enough to detect the brightest 
(10$^{38}$ ergs/s) outbursts from reasonably isolated SMC sources 
such as SMC X-2 \cite{C01}.
One more sensitive ASM that may have potential for the long term study
of the SMC pulsars is the Lobster All Sky Monitor 
that will be installed onboard the International Space Station.
\cite{Fraser02}
This instrument should be very sensitive but source count rates will
be low. Additionally, lobster-type instruments are not sensitive
to harder X-rays.
The ideal type of instrument with which to conduct long term
SMC X-ray pulsar studies may well be an
imaging instrument of several degrees FOV such as on
the proposed MIRAX mission \cite{Staubert03}, continuously 
staring at SMC.

\begin{figure}
  \includegraphics[height=.45\textheight]{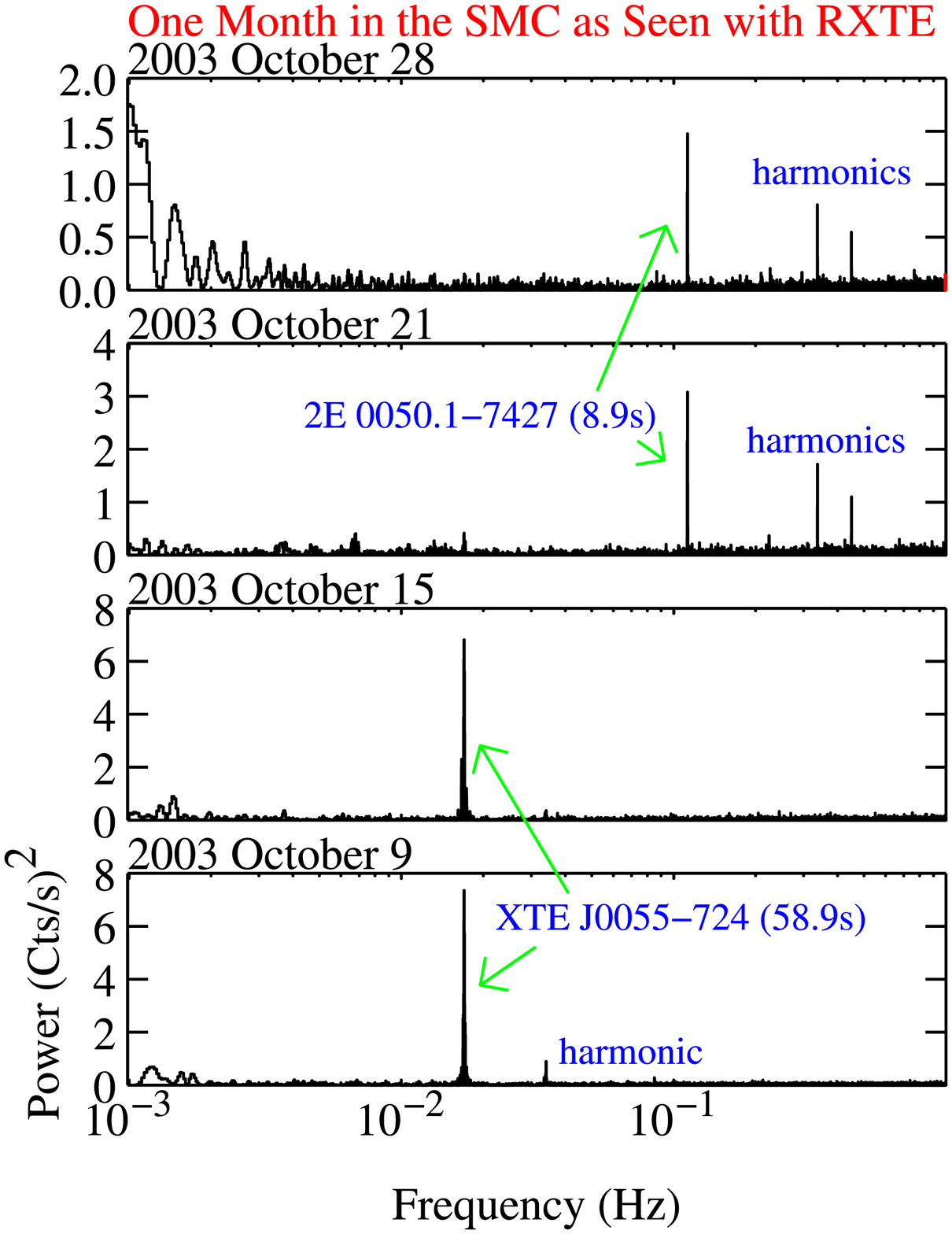}
  \caption{Examples of one recent month
of RXTE observations of the SMC. A power spectrum is shown for each
weekly observation in this month and pulsar detections are marked.
}
\end{figure}

\begin{figure}
  \includegraphics[height=.45\textheight]{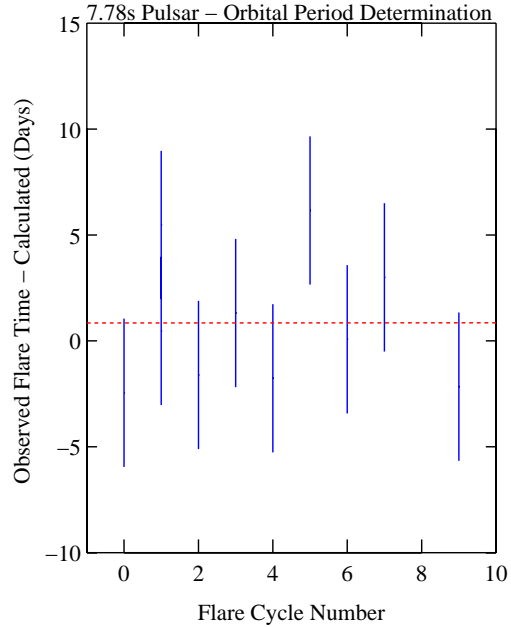}
  \caption{Determination of the orbital period of the 7.78s
pulsar recently identified as SMC X-3. \cite{Edge04b}
}
\end{figure}

\begin{figure}
  \includegraphics[height=.45\textheight]{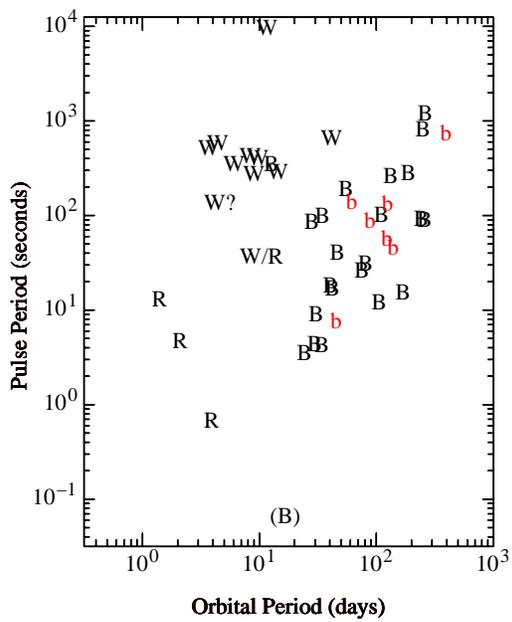}
  \caption{Orbital and spin periods of high-mass X-ray binaries.
``R'' = Roche-lobe overflow mass transfer, ``W'' = wind accretion
system, ``B'' = Galactic Be star source, ``b'' = SMC Be star source,
``(B)'' = LMC Be star source.}
\end{figure}

\begin{figure}
  \includegraphics[height=.45\textheight]{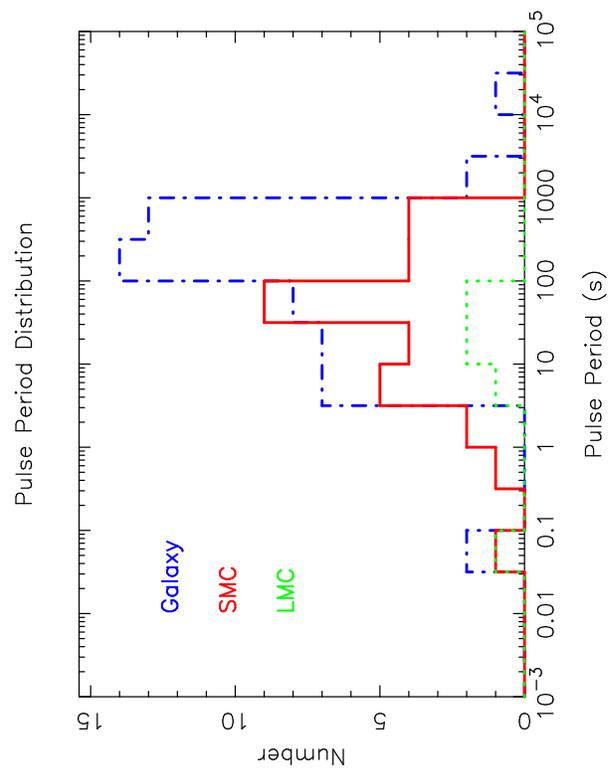}
  \caption{A comparison of the pulse period distributions of the
SMC, Galaxy, and the LMC \cite{L04}. Note that that distributions are {\em not}
normalized but show absolute numbers of pulsars. The SMC distribution
does not contain several recently discovered X-ray pulsars.
}
\end{figure}

%%%%%%%%%%%%%%%%%%%%%%%%%%%%%%%%%%%%%%%%%%%%
%% SAMPLE TABLE
%%
%% Shows the use of \tablehead and \tablenote
%% macros
%%%%%%%%%%%%%%%%%%%%%%%%%%%%%%%%%%%%%%%%%%%%

\begin{table}
\begin{tabular}{rrr}
\hline
    \tablehead{1}{r}{b}{Pulse Period}
  & \tablehead{1}{r}{b}{Orbital Period}  
  & \tablehead{1}{r}{b}{Notes}   \\
\hline
0.716	&	3.9 &	SMC X-1 \\
2.165   & &		XTE J0119-731 \\
2.374	& &		SMC X-2 \\
2.7632	& &		RX J0059.2-7138 \\
3.34	& &		AX J0105-722 \\
4.782   & &      	XTE J0052-723 \\
6.848	& & \\
7.78	& 45.4	& XTE J0055-725 = SMC X-3 \\
8.02    & & \\
8.88	& & 2E 0050.1-7247 \\
9.13	& & 		AX J0049-732 \\
15.3	& &		RX J0052.1-7319 \\
16.6	& &  	\\
18.36	& & \\
22.07	& &		RX J0117.6-7330 \\
31.0	& &		XTE J0111.2-7317 \\
34.08   & & \\
46.4	& & Poor position, not same as 46.6s source? \\
46.6 	& 139 &	XTE, 1WGA J0053.8-7226 \\
51	& &                Same as 25.5s \\
58.97	& 123 &		XTE J0055-724 \\
74.7	& &		AX J0049-729 \\
82.4	& & \\
89      & & \\
91.1	& 88.25	& AX J0051-722 \\
95.2    & &                      \\
101.4	& & AX J0057.4-7325 \\
138.0   & 125 &	Optical period \\
140.1  & &              XMMU J005605.2-722200 \\
144	& 61 &	\\
152.1   & & \\
164.7   & & \\
169.3	& &			XTE J0054-720 \\
172.4	&  & \\
202	& &		XMMU J005920.8-722316 \\
263.6   & & XMMU J004723.7-731226 \\
280.4	& & \\
304.49	& & \\
323	& & RX J0050.8-7316 \\
348/343	& & rapid spin down \\
455+-2  & &      	RX J0101.3-7211 \\ 
503.5   & & \\
564.83	& & \\
701  & & \\
755.5	& 394 &	RX J0049.7-7323 \\

\hline
\end{tabular}
\caption{X-ray Pulsars in the SMC}
\label{tab:a}
\end{table}

%% If the bibliography is produced without BibTeX comment out the

\bibliographystyle{aipproc}   % if natbib is available

\bibliography{corbetr}

\IfFileExists{\jobname.bbl}{}
 {\typeout{}
  \typeout{******************************************}
  \typeout{** Please run "bibtex \jobname" to obtain}
  \typeout{** the bibliography and then re-run LaTeX}
  \typeout{** twice to fix the references!}
  \typeout{******************************************}
  \typeout{}
 }

\end{document}